\documentstyle[12pt]{article}

\begin{document}

\phantom{n}\hfill{January 31, 2000}

\begin{center}
{\bf Evidence for a New State of Matter: \\
An Assessment of the Results from the CERN Lead Beam Programme}

Ulrich Heinz and Maurice Jacob\\
Theoretical Physics Division, CERN, CH-1211 Geneva 23, Switzerland

\end{center}

The year 1994 marked the beginning of the CERN lead beam programme. 
A beam of 33 TeV (or 160 GeV per nucleon) lead ions from the SPS now 
extends the CERN relativistic heavy ion programme, started in the mid 
eighties, to the heaviest naturally occurring nuclei. A run with lead 
beam of 40 GeV per nucleon in fall of 1999 complemented the program
towards lower energies. Seven large experiments participate in the lead 
beam program, measuring many different aspects of lead-lead and lead-gold 
collision events: NA44, NA45/CERES, NA49, NA50, NA52/NEWMASS, 
WA97/NA57, and WA98. Some of these experiments use multipurpose 
detectors to measure simultaneously and correlate several of the 
more abundant observables. Others are dedicated experiments to 
detect rare signatures with high statistics. This coordinated effort 
using several complementing experiments has proven very successful. 
The present document summarizes the most important results from this 
program at the dawn of the RHIC era: soon the relativistic heavy ion 
collider at BNL will allow to study gold-gold collisions at 10 times 
higher collision energies.

Physicists have long thought that a new state of matter could be 
reached if the short range repulsive forces between nucleons could 
be overcome and if squeezed nucleons would merge into one another.
Present theoretical ideas provide a more precise picture for this new
state of matter: it should be a quark-gluon plasma (QGP), in which 
quarks and gluons, the fundamental constituents of matter, are no 
longer confined within the dimensions of the nucleon, but free to move 
around over a volume in which a high enough temperature and/or density 
prevails. This plasma also exhibits the so-called ``chiral symmetry'' 
which in normal nuclear matter is spontaneously broken, resulting in 
effective quark masses which are much larger than the actual masses. 
For the transition temperature to this new state, lattice QCD 
calculations give values between 140 and 180 MeV, corresponding to an 
energy density in the neighborhood of 1 GeV/fm$^3$, or seven times 
that of nuclear matter. Temperatures and energy densities above these 
values existed in the early universe during the first few microseconds 
after the Big Bang. 

It has been expected that in high energy collisions between heavy 
nuclei sufficiently high energy densities could be reached such that 
this new state of matter would be formed. Quarks and gluons would 
then freely roam within the volume of the fireball created by the 
collision. The individual quark and gluon energies would be typical
of a system at very high temperature (above 200 MeV) even if the 
system should not have enough time to fully thermalize. Positive 
identification of the quark-gluon plasma state in relativistic heavy 
ion collisions is, however, extremely difficult. If created, the QGP 
state would have only a very transient existence. Due to color 
confinement, a well-known property of strong interactions at low
energies, single quarks and gluons cannot escape from the collision 
-- they must always combine to color-neutral hadrons before being able 
to travel to the detector. This process is called ``hadronization''. 
Thus, regardless of whether or not QGP is formed in the initial stage, 
the collision fireball later turns into a system of hadrons. In a 
head-on lead-lead collision at the SPS about 2500 particles are 
created (NA49) of which more than 99.9\% are hadrons. Evidence for 
or against formation of an initial state of deconfined quarks and 
gluons at the SPS thus must be extracted from a careful and 
quantitative analysis of the observed final state.
 
{\bf A common assessment of the collected data leads us to conclude that 
we now have compelling evidence that a new state of matter has indeed 
been created, at energy densities which had never been reached over 
appreciable volumes in laboratory experiments before and which 
exceed by more than a factor 20 that of normal nuclear matter. 
The new state of matter found in heavy ion collisions at the SPS 
features many of the characteristics of the theoretically predicted 
quark-gluon plasma.} 

The evidence for this new state of matter is based on a multitude 
of different observations. Many hadronic observables show a strong 
nonlinear dependence on the number of nucleons which participate in 
the collision. Models based on hadronic interaction mechanisms have 
consistently failed to simultaneously explain the wealth of accumulated 
data. On the other hand, the data exhibit many of the predicted 
signatures for a quark-gluon plasma. Even if a full characterization of 
the initial collision stage is presently not yet possible, the data 
provide strong evidence that it consists of deconfined quarks and 
gluons. 

We emphasize that the evidence collected so far is ``indirect'' since it 
stems from the measurement of particles which have undergone significant 
reinteractions between the early collision stages and their final 
observation. Still, they retain enough memory of the initial quark-gluon 
state to provide evidence for its formation, like the grin of the Cheshire 
Cat in Alice in Wonderland which remains even after the cat has disappeared. 
It is expected that the present ``proof by circumstantial evidence'' for 
the existence of a quark-gluon plasma in high energy heavy ion collisions 
will be further substantiated by more direct measurements (e.g. 
electromagnetic signals which are emitted directly from the quarks in 
the QGP) which will become possible at the much higher collision energies 
and fireball temperatures provided by RHIC at Brookhaven and later the
LHC at CERN. 

In the following the most important experimental findings and their
interpretation are described in more detail:

Hadrons are strongly interacting particles. In nuclear collisions, after 
being first created, they undergo many secondary interactions before 
escaping from the collision ``fireball''. When they are finally set free,
the fireball volume has expanded by about a factor 30--50; this 
information can be extracted from two-particle correlations between 
identical hadrons by a method called ``Bose-Einstein interferometry'' 
(NA44, NA49, WA98). At this point, the relative abundances and momentum 
distributions of the hadrons still contain important memories of the 
dense early collision stage which can be extracted by a comprehensive 
analysis of the hadronic final state. More than 20 different hadron 
species, including a few small anti-nuclei (anti-deuteron, anti-helium),
have been measured by the seven experiments (NA44, NA45, NA49, NA50, 
NA52, WA97, WA98). A combined analysis of their momentum distributions
and two-particle correlations shows that, at the point where they
stop interacting and ``freeze out'', the fireball is in a state of
tremendous explosion, with expansion velocities exceeding half the 
speed of light, and very close to local thermal equilibrium at a 
temperature of about 100-120 MeV. This characteristic feature gave 
rise to the name ``Little Bang''. The observed explosion calls for 
strong pressure in the earlier collision stages. Recently 
measured anisotropies in the angular distribution of the momenta 
perpendicular to the beam direction (NA49, NA45, WA98) indicate that 
the pressure was built up quickly, pointing to intense rescattering 
in the early collision stages.

An earlier glimpse of the expanding system is provided by a measurement 
of correlated electron-positron pairs, also called dileptons (NA45). 
These data show that in sulphur-gold and lead-gold collisions the 
expected peak from the rho ($\rho$) vector meson (a particle which 
can decay into dileptons even before freeze-out) is completely smeared 
out. Simultaneously, NA45 finds in lead-gold collisions an excess of 
dileptons in the mass region between 250 and 700 MeV, by about a factor 
3 above expectations from hadron decays scaled from proton-nucleon to 
lead-gold collisions. Theory explains this by a broadening of the 
$\rho$'s spectral function, resulting from scattering among pions and 
nucleons in a very dense hadronic fireball, just below the critical 
energy density for quark-gluon plasma formation. The $\rho$ meson 
mixes with its partner under chiral symmetry transformations, signalling 
the onset of chiral symmetry restoration as matter becomes denser and 
denser.

The theoretical analysis of the measured hadron abundances (NA44, 
NA45, NA49, NA50, NA52, WA97, WA98) shows that they reflect a state 
of ``chemical equilibrium'' at a temperature of about 170 MeV. This 
points to an even earlier stage of the collision. In fact, such 
temperatures (corresponding to an energy density of about 1 GeV/fm$^3$) 
are the highest allowed ones before, according to lattice QCD,
hadrons should dissolve into quarks and gluons. The observations are 
explained by assuming that at this temperature the hadrons were formed 
by a statistical hadronization process from a pre-existing quark-gluon 
system. Theoretical studies showed that at CERN energies subsequent 
interactions among the hadrons, while causing pressure and driving the 
expansion and cooling of the fireball, are very ineffective in changing 
the abundance ratios. This is why, after accounting for the decay of 
unstable resonances, the finally measured hadron yields reflects rather 
accurately the conditions at the quark-hadron transition.

A particularly striking aspect of this apparent ``chemical 
equilibrium'' at the quark-hadron transition temperature is the
observed enhancement, relative to pro\-ton-induced collisions, of 
hadrons containing strange quarks. Globally, when normalized to the 
number of participating nucleons, this enhancement corresponds to a 
factor 2 (NA49), but hadrons containing more than one strange quark 
are enhanced much more strongly (WA97, NA49, NA50), up to a factor 15 
for the Omega ($\Omega$) hyperon and its antiparticle (WA97)! Lead-lead 
collisions are thus qualitatively different from a superposition of
independent nucleon-nucleon collisions. That the relative enhancement 
is found to {\em increase} with the strange quark content of the 
produced hadrons contradicts predictions from hadronic rescattering
models where secondary production of multi-strange (anti)baryons is 
hindered by high mass thresholds and low cross sections. Since the 
hadron abundances appear to be frozen in at the point of hadron 
formation, this enhancement signals a new and faster strangeness-producing 
process {\em before or during} hadronization, involving intense rescattering 
among quarks and gluons. This effect was predicted about 20 years 
ago as a quark-gluon plasma signature, resulting from a combination
of large gluon densities and a small strange quark mass in this 
color deconfined, chirally symmetric state. Experimentally it is 
found not only in lead-lead collisions, but even in central 
sulphur-nucleus collisions, with target nuclei ranging from sulphur 
to lead (NA35, WA85, WA94). This is consistent with estimates of 
initial energy densities above the critical value of 1 GeV/fm$^3$ 
even in those collisions. 

Evidence for the formation of a transient quark-gluon phase without 
color confinement is further provided by the observed suppression 
of the charmonium states $J/\psi,\,\chi_c,$ and $\psi'$ (NA50). These 
particles contain charmed quarks and antiquarks ($c$ and $\bar c$)
which are so heavy that they can only be produced at the very beginning 
when the constituents of the colliding nuclei still have their full 
energy. As one varies the size of the colliding nuclei and the 
centrality of the collision one finds, after subtracting the 
expected absorption effects from final state interactions between 
the $c\bar c$ pair and the nucleons of the interpenetrating nuclei,  
a succession of suppression patterns: The most weakly bound 
state, $\psi'$, is suppressed already in sulphur-uranium collisions 
(NA38), the intermediate $\chi_c$ seems to disappear quite suddenly 
in semicentral lead-lead collisions, and in the most central lead-lead 
collisions an additional reduction of the $J/\psi$ yield indicates that 
now also the strongly bound $J/\psi$ ground state itself is 
significantly suppressed (NA50). The observation of $\chi_c$ suppression 
is indirect, via its 30-40\% contribution to the measured $J/\psi$ 
yield which is expected from scaling proton-proton measurements. 
Charmonium suppression was predicted 15 years ago as a consequence of 
color screening in a quark-gluon plasma which should keep the charmed 
quark-antiquark pairs from binding to each other. According to this 
prediction, suppressing the $J/\psi$ requires temperatures which are 
about 30\% above the color deconfinement temperature, or energy 
densities of about 3 GeV/fm$^3$. This agrees with estimates of the 
initial energy densities reached in central lead-lead collisions, 
based on calorimetry or on a back-extrapolation from the freeze-out 
stage to the timei before expansion started. It was tried to reproduce 
the data by assuming that the charmonia are destroyed solely by final 
state interactions with surrounding hadrons; none of these attempts 
can account for the shape of the centrality dependence of the observed 
suppression. On the other hand, the interpretation of this pattern 
in terms of color screening by deconfined quarks and gluons 
leads to the prediction of a similar suppression pattern at RHIC in 
much smaller nuclei; this prediction will soon be tested.

{\sf In spite of its many facets the resulting picture is simple: the two
colliding nuclei deposit energy into the reaction zone which materializes
in the form of quarks and gluons which strongly interact with each other.
This early, very dense state (energy density about 3--4 GeV/fm$^3$, mean 
particle momenta corresponding to $T$\,$\approx$\,240 MeV) suppresses the 
formation of charmonia, enhances strangeness and begins to drive the 
expansion of the fireball. Subsequently, the ``plasma'' cools down and 
becomes more dilute. At an energy density of 1~GeV/fm$^3$ 
($T$\,$\approx$\,170 MeV) the quarks and gluons hadronize and the final 
hadron abundances are fixed. At an energy density of order 50 MeV/fm$^3$ 
($T$\,=\,100--120 MeV) the hadrons stop interacting, and the fireball 
freezes out. At this point it expands with more than half the light 
velocity.}

This does not happen only in a few ``special'' collision events, but 
essentially in {\em every} lead-lead collision: characteristic 
observables, like the average transverse momentum of produced particles
or the kaon/pion ratio, show only the statistically expected fluctuations
in a thermalized ensemble, around average values which are the same in 
all collisions (NA49). Since the kaon/pion ratio is essentially fixed at 
the point of hadronization, this indicates the absence of long-range
correlations like those expected in a fully-developed thermodynamic 
phase transition. A better theoretical understanding of the 
phase-transition dynamics might emerge from these observations. The 
short-range character suggests similarities with the transition 
found in high-$T_c$ superconductivity.

``Direct'' observation of the quark-gluon plasma may be possible via
electromagnetic radiation emitted by the quarks during the hot initial 
stage. Searches for this radiation were performed at the SPS (WA98, 
NA45, NA50) but are difficult due to high backgrounds from other 
sources. For sulphur-gold collisions WA80 and NA45 established that 
not more than 5\% of the observed photons are emitted directly. For
lead-lead collisions WA98 have reported indications for a significant 
direct photon contribution. Preliminary data from NA45 are consistent 
with this finding, but so far not statistically significant. NA50 has 
seen an excess by about a factor 2 in the dimuon spectrum in the mass 
region between the $\phi$ and $J/\psi$ vector mesons. The predicted 
electromagnetic radiation rates at the above mentioned temperatures 
are marginal for detection. While under these conditions it is a great 
experimental achievement to have obtained positive evidence for a signal, 
its connection with the predicted ``thermal plasma radiation'' is not yet 
firmly established. 

This is expected to change at the higher collision energies provided 
by RHIC and LHC. The much higher initial temperatures (up to nearly 
1000~MeV for lead-lead collisions at the LHC have been predicted) and 
longer plasma lifetimes should facilitate the direct observation of 
the plasma radiation and lead to the production of additional heavy 
charm quarks by gluon-gluon scattering in the QGP phase. The much 
higher initial energy densities which can be reached at RHIC and LHC 
give us more time until the quarks and gluons rehadronize, thus 
allowing for a quantitative characterization of the quark-gluon plasma 
and detailed studies of its early tharmalization processes and dynamical 
evolution. Finally, the higher collision energies allow for the 
production of jets with large transverse momenta, whose leading quarks 
can be used as ``hard penetrating probes'' within the quark-gluon plasma. 
At RHIC a set of four large detectors, with complementary goals and 
capabilities, ensures that all experimental aspects of 
ultrarelativistic heavy ion collisions are optimally covered.
The ability of the collider to simultaneously accelerate and collide 
nuclei of different sizes and energies promises a complete understanding
of systematic trends as one proceeds from proton-proton via proton-nucleus
to gold-gold collisions. As in solid state physics, where the knowledge
of the basic interaction Lagrangian (QED) does not permit to reliably
predict many bulk properties and where the detailed understanding of 
the latter is usually driven by experiment, we expect that such a 
systematic experimental study of strongly interacting matter will 
eventually lead to a quantitative understanding of ``bulk QCD''. We 
are looking forward to these far-reaching opportunities provided by 
RHIC and LHC.
  
\bigskip

\noindent{\bf Key references to the experimental data:} 

\parskip 0.2cm

\noindent
{\sf NA44 Collaboration:}

\noindent
H. Beker et al., ``$M_T$-dependence of boson interferometry in heavy ion
collisions at the CERN SPS'', Physical Review Letters {\bf 74} 
(1995) 3340-3343

\noindent
I.G. Bearden et al., ``Collective expansion in high-energy heavy ion 
collisions'', Physical Review Letters {\bf 78} (1997) 2080-2083

\noindent
I.G. Bearden et al., ``Strange meson enhancement in Pb-Pb collisions'',
Physics Letters B {\bf 471} (1999) 6-12

\noindent
{\sf NA45/CERES Collaboration:}

\noindent
G. Agakichiev et al., ``Low-mass $e^+e^-$ pair production in 158 A GeV 
Pb-Au collisions at the CERN SPS, its dependence on multiplicity and
transverse momentum'', Physics Letters B {\bf 422} (1998) 405-412

\noindent
B. Lenkeit et al., ``New results on low-mass lepton pair production in 
Pb-Au collisions at 158 GeV/$c$ per nucleon'', Nuclear Physics A {\bf 654} 
(1999) 627c-630c

\noindent
B. Lenkeit et al., ``Recent results from Pb-Au collisions at 158 GeV/$c$ 
per nucleon obtained with the CERES spectrometer'', Nuclear Physics
A {\bf 661} (1999) 23c-32c 

\noindent
{\sf NA49 Collaboration:}

\noindent
T. Alber et al., ``Transverse energy production in $^{208}$Pb+Pb collisions 
at 158 GeV per nucleon'', Physical Review Letters {\bf 75} (1995) 3814-3817

\noindent
H. Appelsh\"auser et al., ``Hadronic expansion dynamics in central Pb+Pb 
collisions at 158 GeV per nucleon'', European Physical Journal C {\bf 2}
(1998) 661-670


\noindent
F. Sikler et al., ``Hadron production in nuclear collisions from the 
NA49 experiment at 158 GeV/$c \cdot A$'', Nuclear Physics A {\bf 661} 
(1999) 45c-54c

\noindent
{\sf NA50 Collaboration:}

\noindent
M.C. Abreu et al., ``Anomalous $J/\psi$ suppression in Pb-Pb interactions
at 158 GeV/$c$ per nucleon'', Physics Letters B {\bf 410} (1997) 337-343

\noindent
M.C. Abreu et al., ``Observation of a threshold effect in the anomalous
$J/\psi$ suppression'', Physics Letters B {\bf 450} (1999) 456-466

\noindent
M.C. Abreu et al., ``Evidence for deconfinement of quarks and gluons
from the $J/\psi$ suppression pattern measured in Pb-Pb collisions at 
the CERN-SPS'', CERN-EP-2000-013, submitted to Physics Letters B 

\noindent
{\sf NA52/NEWMASS Collaboration:}

\noindent
R. Klingenberg et al., ``Strangelet search and antinuclei production 
studies in Pb+Pb collisions'', Nuclear Physics A {\bf 610} (1996) 306c-316c
 
\noindent
G. Ambrosini et al., ``Baryon and antibaryon production in Pb-Pb 
collisions at 158 $A$ GeV/$c$'', Physics Letters B {\bf 417} (1998) 202-210

\noindent
G. Ambrosini et al., ``Impact parameter dependence of $K^{\pm}, p, 
\overline{p}, d$ and $\overline{d}$ production in fixed target Pb + Pb 
collisions at 158 GeV per nucleon'', New Journal of Physics {\bf 1}
(1999) 22.1-22.23

\noindent
{\sf WA97/NA57 Collaborations:}

\noindent
E. Andersen et al., ``Strangeness enhancement at mid-rapidity in Pb-Pb 
collisions at 158 A GeV/c'', Physics Letters B {\bf 449} (1999) 401-406

\noindent
F. Antinori et al., ``Production of strange and multistrange hadrons in 
nucleus-nucleus collisions at the SPS'', Nuclear Physics A {\bf 661} (1999) 
130c-139c

\noindent
F. Antinori et al., ``Transverse mass spectra of strange and multistrange 
particles in Pb-Pb collisions at 158 $A$ GeV/$c$'', CERN-EP-2000-001, 
submitted to European Physical Journal C

\noindent
{\sf WA98 Collaboration:}

\noindent
R. Albrecht et al., ``Limits on the production of direct photons in 200 $A$ 
GeV $^{32}$S+Au collisions'', Physical Review Letters {\bf 76} (1996) 
3506-3509

\noindent
M.M. Aggarwal et al., ``Centrality dependence of neutral pion production 
in 158 $A$ GeV $^{208}$Pb+$^{208}$Pb collisions, Physical Review Letters
{\bf 81} (1998) 4087-4091; {\bf 84} (2000) 578-579(E)

\noindent
M.M. Aggarwal et al., ``Freeze-out parameters in central 158 $A$ GeV 
$^{208}$Pb+$^{208}$Pb collisions'', Physical Review Letters {\bf 83}
(1999) 926-930 

\end{document}